\documentclass{evn2004}
\usepackage{txfonts}
\usepackage{graphicx}
\begin{document}
   \title{Multi-Frequency VLBI Observations of GHz-Peaked Spectrum Sources}

   \author{S. Kameno\inst{1}, M. Inoue\inst{1}, Z.-Q. Shen\inst{2}, S. Sawada-Satoh\inst{3},
          \and
          K. Wajima\inst{4}
          }
	\authorrunning{S. Kameno {\it et al.}}
	\titlerunning{Multi-Frequency VLBI observations of GPS sources}

   \institute{National Astronomical Observatory of Japan, 2-21-1 Osawa, Mitaka, Tokyo 181-8588 Japan
   \and
   Shanghai Astronomical Observatory, 80 Nandan Road, Shanghai  200030, China
   \and
   Academia Sinica Institute of Astronomy and Astrophysics, P.O. Box 23-141, Taipei 106, Taiwan
   \and
	Korea Astronomy Observatory, 61-1 Hwaam-dong, Yuseong, Daejeon 305-348, Korea
   }

   \abstract{
We report results of pentachromatic VLBI survey for 18 GHz-peaked spectrum sources. Spectral fitting at every pixel across five frequencies allows us to illustrate distribution of optical depth in terms of free--free absorption or synchrotron self absorption. Quasars and Seyfert 1 sources show one-sided morphology with a core at the end where the optical depth peaks. Radio galaxies and Seyfert 2 show symmetric double-sided jets with a optically thick core at the center.
   }

   \maketitle
%

\section{Introduction}
GHz-Peaked Spectrum (GPS) sources are powerful radio sources associated with active galactic nuclei, which show a convex radio spectrum peaked at GHz frequencies and a compact overall size smaller than kpc dimension (O'Dea, Baum, \& Stanghellini \cite{1991ApJ...380...66O}; O'Dea \cite{1998PASP..110..493O}).

A power-law spectrum at frequencies higher than the spectral peak is produced by optically thin synchrotron radiation. 
A spectral cutoff indicates that the synchrotron emission is optically thick at lower frequencies than the peak, in terms of synchrotron self absorption (SSA) or free--free absorption (FFA).

O'Dea \& Baum (\cite{1997AJ....113..148O}) and Snellen et al. (\cite{2000MNRAS.319..445S}) showed correlation between the peak frequency and the overall linear size, and claimed that SSA controls the the peak frequency.
On the other hand, Bicknell, Dopita, \& O'Dea (\cite{1997ApJ...485..112B}) pointed out that FFA through ionized gas surrounding radio lobes can also produce the correlation between the peak frequency and the overall size. 
Evidence for FFA was found via multi-frequency VLBI observations in particular GPS sources (e.g. NGC 1052; Kameno et al. \cite{2001PASJ...53..169K}; Vermeulen et al. \cite{2003A&A...401..113V}; Kameno et al. \cite{2003PASA...20..134K}; Kadler et al. \cite{2004_to_appear_in_A&A}).

A power-law spectrum of synchrotron emitter through FFA will be affected as
\begin{eqnarray}
S_{\nu} = S_0 \nu^{\alpha} \exp(-\tau_0 \nu^{-2.1}), \label{eqn:ffaspec}
\end{eqnarray}
where $\tau_0$ is the optical depth at 1 GHz.
$\tau_0$ is related with the electron density $n_{\rm e}$ and temperature $T_{\rm e}$ as $\tau_0 = \int_{\rm LOS} 0.46 n_{\rm e}^2 T_{\rm e}^{-3/2} dL$.
Thus, multi-frequency VLBI observations enable spectral fitting at every pixel to measure $\tau_0$ and illustrate the distribution of cold dense plasma.

\section{The Survey}

We have carried out five-frequency VLBI observations for 18 GPS sources.
The VSOP was used at 1.6 and 4.8 GHz to obtain comparable beam sizes with those at 2.3, 8.4, and 15.4 GHz with the VLBA. \\
Detailed image performance is shown in the URL {\tt http://vsop.mtk.nao.ac.jp/\%7Ekameno/GPSsurvey/}.

Table \ref{tab:sourcelist} lists the sample sources which were selected from the GPS catalog by de Vries, Barthel, \& O'Dea (\cite{1997A&A...321..105D}).
Selection criteria of (1) peak frequency $\nu_{\rm m}$ stand within $1.6$ GHz $< \nu_{\rm m} < 15$ GHz, (2) flux density limitation of $S_{\rm 1.6 GHz} > 0.1$ Jy, $S_{\rm 5 GHz} > 0.5$ Jy, and $S_{\rm 15 GHz} > 0.2$ Jy, are applied.
Five radio galaxies and 13 quasars are included in the sample.

   \begin{figure}
	\vspace{50mm}
	\includegraphics{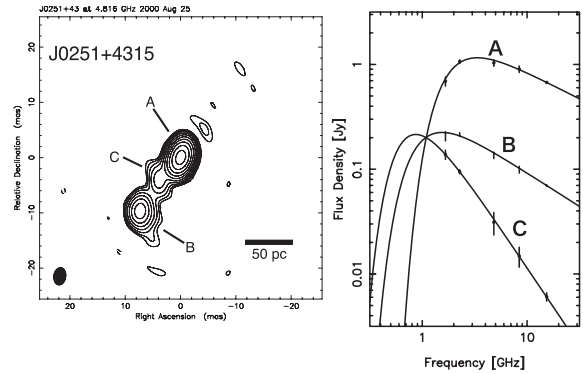}
      \caption{Morphology and spectra of a type-1 GPS source J0251+4315 .
         \label{fig:Type1sample}
         }
   \end{figure}
   \begin{figure}
	\vspace{30mm}
	\includegraphics{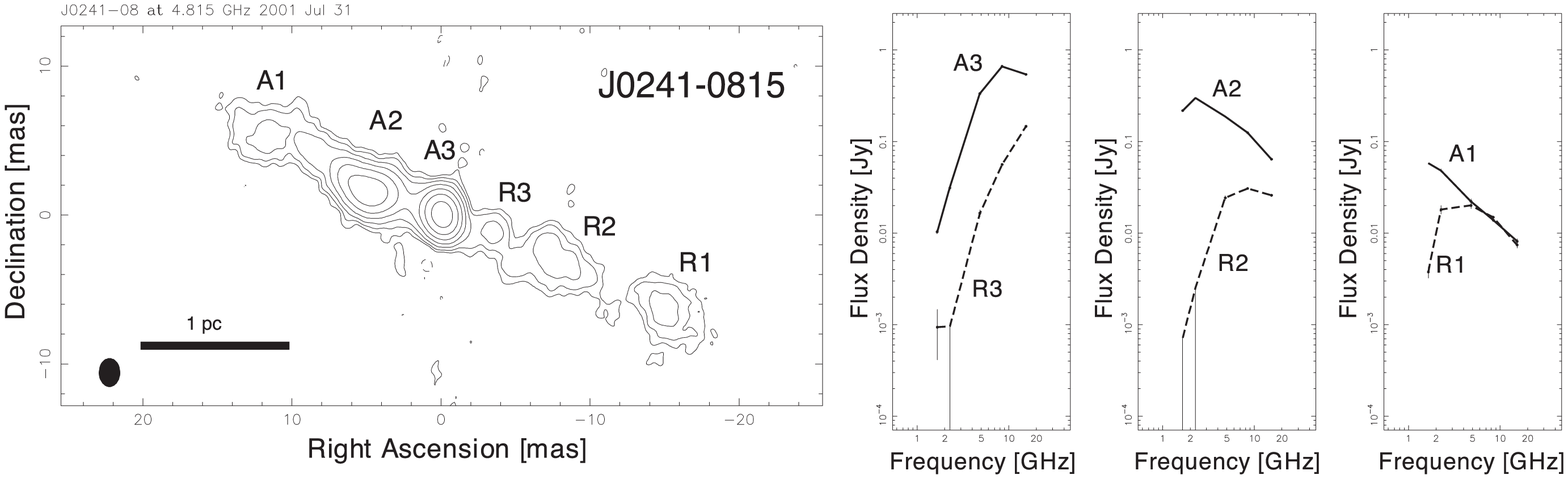}
      \caption{Morphology and spectra of a type-2 GPS source J0241$-$0815.
         \label{fig:Type2sample}
         }
   \end{figure}
%
   \begin{table}
      \caption[]{GPS sample sources.}
         \label{tab:sourcelist}
\begin{tabular}{lllcl} \hline 
\multicolumn{1}{c}{J2000} &
\multicolumn{1}{c}{other} &
\multicolumn{1}{c}{$z$} &
\multicolumn{1}{c}{class} &
\multicolumn{1}{c}{$\alpha$} \\ \hline
J0111+3906 & 0108+388 &
	$0.66847$ &
	2 &
	$-1.1$ \\
J0203+1134 & 0201+113 &
	$3.6109$ &
	1 &
	$-0.3$ \\
J0241-0815 & NGC 1052 &
	$0.0049$ &
	2 &
	$-1.0$ \\
J0251+4315 & 0248+430 &
	$1.31$ &
	1 &
	$-1.0$ \\
J0503+0203 & 0500+019 &
	$0.58457$ &
	2 &
	$-1.0$ \\
J0650+6001 & 0646+600 &
	$0.455$ &
	1 &
	$-0.6$ \\
J0741+3112 & 0738+313 &
	$0.635$ &
	1 &
	$-0.5$ \\
J0745+1011 & 0742+103 &
	$-$ &
	1 &
	$-0.7$ \\
J0745-0044 & 0743-006 &
	$0.994$ &
	1 &
	$-1.0$ \\
J0905+4850 & 0902+490 &
	$2.690$ &
	1 &
	$-0.7$ \\
J1146-2447 & 1143-245 &
	$-$ &
	1 &
	$-0.7$ \\
J1335+4542 & 1333+459 &
	$2.449$ &
	1 &
	$-1.2$ \\
J1845+3541 & 1843+356 &
	$0.746$ &
	2 &
	$-1.0$ \\
J1850+2825 & 1848+283 &
	$2.56$ &
	1 &
	$-0.7$ \\
J2052+3635 & 2050+364 &
	$0.354$ &
	2 &
	$-1.5$ \\
J2129-1538 & 2126-158 &
	$3.268$ &
	1 &
	$-0.7$ \\
J2151+0552 & 2149+056 &
	$0.74$ &
	1 &
	$-0.75$ \\
J2340+2641 & 2337+264 &
	$-$ &
	1 &
	$-0.7$ \\ \hline
\end{tabular}

{\it Columns}: (1) IAU source name; (2) other source name; (3) redshift; (4) classification. type-1 stands for quasars and Seyfert 1, where type-2 for radio galaxies and Seyfert 2; (5) spectral index at frequencies above the peak.
   \end{table}
%
%

   \begin{figure}
	\vspace{115mm}
	\includegraphics{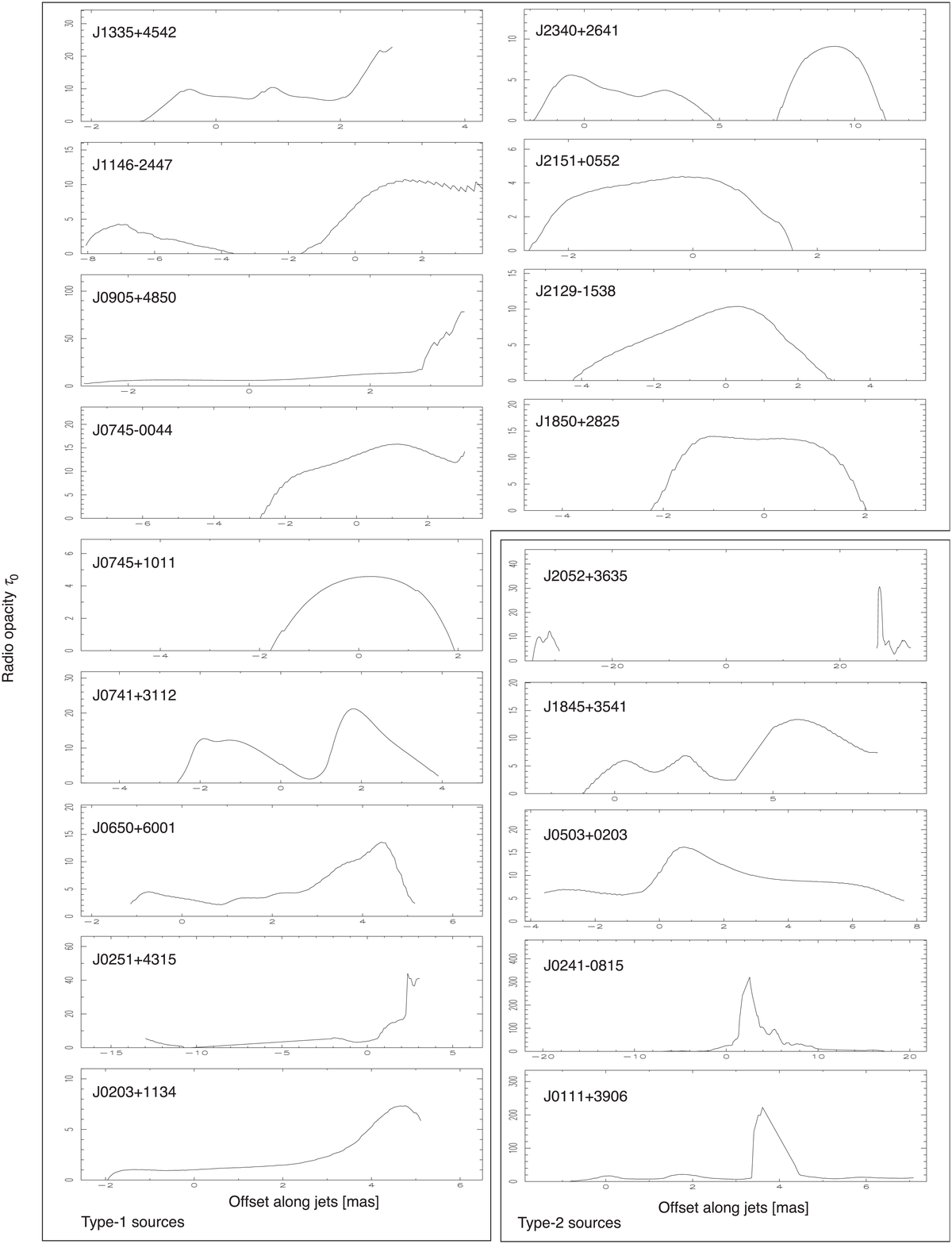}
      \caption{Opacity profile along jets in GPS sources.
         \label{fig:Opacity}
         }
   \end{figure}

Images at five frequencies are restored with a common Gaussian beam whose FWHM is a geometric mean of five synthesized beams.
These images are registered by referencing several distinct components.
Then, spectral fitting was applied for every pixel across five frequencies using equation \ref{eqn:ffaspec}.

Figure 1 and 2 show total intensity maps and results of spectral fit for archetypes of type-1 (J0251+4315) and -2 (J0241$-$0815).
Figure 3 shows profiles of $\tau_0$ along the jet.

Most of type-1 sources (J0203+1134, J0251+4315, J0650+6001, J0741+3112, J0745-0044, J0905+4850, J1146-2447, J1335+4542, J2129-1538, J2151+0552, J2340+2641) show core-jet structure, though J0745+1011 and J1850+2825 are almost unresolved.
The optical depth tends to peak at the core component, which is unresolved and brightest at 15.4 GHz.
It decreases downstream along the jet, and enhances at knots or hot spots.
This behavior can be understood in the context of synchrotron self absorption.

Opacity profiles in type-2 sources tends to be symmetry along the jets.
The optical depth peaks at the central component where the nucleus is considered to be, and decreases both downstream.
Spiky profiles at the central component of J0111+3906 and J0241$-$0815 mark the maximum optical depth of $\tau_0 > 100$, which requires $n_{\rm e}^2 L > 10^8$ cm$^{-6}$ pc.
This profile and extreme column density can be understood by an edge-on plasma torus perpendicular to the jet.
J0503+0203 and J1845+3541 shows relatively asymmetric opacity profiles with lower maximum.
Slant non-edge-on viewing angle can produce such asymmetry.
J2052+3635, the largest object in our sample, show a double lobe with multiple hot spots.
A central component, which is slightly visible at only 4.8 GHz, is too weak to measure the optical depth.

\section{Conclusions}
We have performed pentachromatic VLBI survey for type-1 and -2 GPS sources to illustrate distribution of optical depth.
Type-1 sources, consisting of quasars and Seyfert 1, show one-sided core-jet structure. The optical depth peaks at the core component, and decreases downstream with enhancements at knots and hot spots.
Type-2 sources, including radio galaxies and Seyfert 2, show symmetric double-sided structure. The optical depth peaks at the central component with a significantly large opacity.

\begin{acknowledgements}
We thank the European community for collaborations to the VSOP space VLBI project. Many VSOP observations owe ground radio telescopes in the European VLBI network.
\end{acknowledgements}

\end{document}